\documentclass[allclo]{FBSart}
\usepackage{amsfonts}
\usepackage{amssymb}
\usepackage{epsfig}

\title{QCD Phenomenology of Static Sources}
\author{Gunnar S. Bali}
\institute{Department of Physics \& Astronomy, University of Glasgow,
Glasgow G12 8QQ, UK}

\runningauthor{Gunnar\,S.\,Bali}
\runningtitle{QCD Phenomenology of Static Sources}
\begin{document}

\maketitle
\begin{abstract}
We discuss the spectrum of open string and point particle excitations
in QCD with various source representations. Some general relations
are introduced and lattice results presented. In particular we discuss
the short-distance behaviour, relate this to perturbation theory expectations
and comment on the matching between low energy matrix elements and
high energy Wilson coefficients, within the framework of effective
field theories.
\end{abstract}

\section{Introduction}
We discuss the excitation spectrum of QCD in its static
limit. This is quite amusing, in the context of a Light Cone Workshop.
However, some of the results are instructive and of a very general nature,
in particular those concerning the spectrum of QCD and the
question of power divergences and renormalons. For instance
the Wilson-Schwinger line appears at prominent places. We shall
see that in a scheme without a hard cut-off, such as dimensional
regularisation (DR), such objects require special attention.
Wilson lines do not
only appear in the static limit, within the framework of heavy quark effective
field theories, but for instance also within light cone parton distributions
\cite{Collins:1981uw}, if one wishes to define them in a gauge
invariant (and hopefully path-independent)
way~\cite{Ji:2002aa,Collins:2003fm,Burkardt:2003uw}.

In general, a Wilson-Schwinger line of length $l$ in Euclidean
space-time within a correlation
function will result in a term $\propto \exp(-\overline{\Lambda}l)$, with some
self-energy $\overline{\Lambda}$. In DR
any perturbative contribution to
$\overline{\Lambda}$ vanishes. This is different in schemes with a
hard momentum cut-off, such as lattice regularisation. In this case
$\overline{\Lambda}$ contains a contribution that is proportional to the
inverse lattice spacing $a^{-1}$ and which can be expanded in powers
of $\alpha_s$.
Within physical observables
such power terms obviously
have to cancel. 

To translate $\overline{\Lambda}$ from the lattice into
the on-shell (OS) scheme
the perturbative expansion of $\overline{\Lambda}$ has to be subtracted,
replacing the $a^{-1}$ term by a renormalon ambiguity.
For physical observables to remain unaffected,
the renormalon of $\overline{\Lambda}$ in
the OS scheme has to cancel against a similar renormalon
associated with a different term.
One nice  thing about a hard cut-off is that the structure
of power terms within such a regularisation
scheme is indicative of the renormalon structure that one
will encounter in a DR calculation. One way of elucidating this
connection has been explained above: the price of cancelling
a power term is a renormalon. Vice versa it is possible
to remove (part of) such a renormalon, by introducing a
scale dependence and replacing it by a power term.
One such continuum scheme, the RS scheme, has been suggested in
refs.~\cite{staticpot,Bali:2003jq}.

Here I discuss the spectrum of QCD in the static limit.
The simplest case would be that of a static-light meson.
In the continuum limit the mass $\overline{\Lambda}$
of such a particle will diverge.
Only level splittings are well defined. One can however,
relate this mass to the mass of the physical $B$ meson, within
Heavy Quark Effective Theory (HQET). In this case the
power term/renormalon of $\overline{\Lambda}$ will be compensated
for by a similar contribution to $m_b$.

This article is organised as follows: 
in Section~\ref{sec:string} we introduce the relevant correlation
functions
and discuss some fundamental relations.
Then in Section~\ref{sec:large} we survey some aspects of the
large distance behaviour and string breaking, before 
we discuss short distance relations in Section~\ref{sec:short}.
\section{Static QCD strings and particles: the broad picture}
\label{sec:string}
We discuss the spectrum of QCD with one or two static charges, where we
restrict ourselves to charges in the fundamental representation,
${\mathbf 3}$, and the adjoint representation, ${\mathbf 8}$.
The former can be motivated as representing the
starting point for realising a heavy
quark within an effective field theory such as HQET, non-relativistic
QCD (NRQCD) or potential NRQCD (pNRQCD)~\cite{Pineda:1997bj,Brambilla:1999xf}.
The
adjoint representation can be used to describe a gluino, 
within a heavy-gluino expansion (should a gluino exist and be sufficiently
stable).

\begin{figure}[hbt]
\epsfig{file=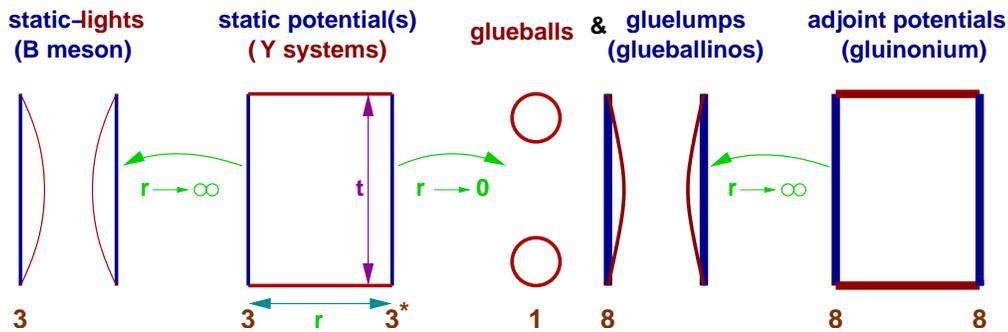,width=.99\columnwidth}
\caption[]{Correlation functions in the static sector of QCD.}\label{fig:objects}
\end{figure}

We schematically depict the correlation functions $C(r,t)$
that we investigate in Fig. \ref{fig:objects}. The states are created
at time $0$ and destroyed at time $t$. Straight spatial lines are
parallel transporters, straight temporal lines static propagators
(Wilson-Schwinger lines) in fundamental or adjoint representation.
Curved lines represent sea quarks and gluons.
The corresponding (in most cases
$r$-dependent) energy levels
can be extracted in the limit of large Euclidean times,
\begin{equation}
E(r)=-\lim_{t\rightarrow\infty}\frac{d\ln C(r,t)}{dt}.
\end{equation}

We denote the energy level of a static-light meson by,
\begin{equation}
M_B=m_b+\overline{\Lambda}+O(1/m_b),
\end{equation}
where both the quark mass $m_b$ and the binding energy (or static energy)
$\overline{\Lambda}$ are scheme- and scale-dependent, in reflection of the
fact that asymptotic states cannot carry strong charge.
$M_B$ however is a physical observable. Note that
$m_b$ does not appear as a scale
in the static limit. I introduce it nonetheless
for pedagogical reasons.
Taking account of light quark spin and angular momentum
the spectrum of $\overline{\Lambda}$
is arranged in spin-doublets $(n+\frac12)^{\pm}$ where for instance
$\frac12^{-}$ corresponds to $l=0$ and $\frac12^{+}$ to $l=1$.
The presence of such spin doublets
has found an exciting application, with the
recent discovery of
two new $D_s$ states~\cite{Nowak:1992um,Bardeen:1993ae,Bali:2003jv,Green:2003zz}. Once $1/m_b$ corrections
are added, the heavy quark spin has to be taken into account and
the $\frac12^{-}$ ground state will further split into a pseudoscalar
$B$ meson and a vector $B^*$ meson. 

The next step is an open string state, between a fundamental source
and anti-source separated by a distance $r$.
In this case the inclusion of $1/m_b$ corrections is
essential, even at leading order as without a kinetic term all sorts
of pathologies can arise which are related to the fact that
heavy quarks at relative speed decouple from each other.
The more conceptional results presented
here are not affected by this problem.
One encounters a ground state $E_s(r)$ as well as hybrid excitations
$E_H(r)$, due to radial and/or orbital gluonic excitations. 
Obviously, sea quarks screen these potentials and,
\begin{equation}
\lim_{r\rightarrow\infty}E_s(r)=2M_B,
\end{equation}
which is known as ``string breaking''.
Note that the heavy quark mass $m_b$
cancels from the combination, $E_s(r)-2M_B$.
A similar relation will apply to $E_H$, which will also decay, depending
on the quantum numbers $H$, into a pair of $B$ mesons.

Within the framework of pNRQCD
which
amounts to a double expansion both in the relative quark 
velocity (NRQCD) and in the distance $r$,
one can classify the short distance behaviour. In the static limit
the only scale, apart from a typical non-perturbative
QCD scale $\Lambda_{QCD}$, is the distance $r$ and one can
identify~\cite{Bali:2003jq},
\begin{eqnarray}
E_s(r)&=&2m_b+V_s(r)+O(r^2)=2m_b^L+E_{s}^L(r),\label{eq:es}\\
E_H(r)&=&2m_b+V_o(r)+\Lambda_H+O(r^2)=2m_b^L+E_{H}^L(r)\label{eq:eo},
\end{eqnarray}
where both, singlet and octet potentials $V_s(r)$ and $V_o(r)$ are
calculable in perturbation theory for distances,
$r\ll\Lambda_{QCD}^{-1}$. $\Lambda_H$ stands for the mass of a gluelump
(see below). Again, $m_b$, $V_s(r)$, $V_o(r)$ and $\Lambda_H$ are
scheme and scale dependent. The new ingredient now is that
the definition of the infrared gluelump energy $\Lambda_H$ carries an
ultraviolet renormalon ambiguity that has to be compensated for by $V_o$.
The ultraviolet ambiguity of $V_o$ in turn is the same as that of $V_s$
and related to the definition of the quark mass $m_b$.
On the other hand one can also calculate these energy levels non-perturbatively
in lattice simulations and identify these as the ground state static
potential,
$E_{s}^L(r)$ (which is in the $s=\Sigma_g^+$ representation of the
cylindrical group $D_{\infty h}$) and hybrid excitations, $E_H^L(r)$.

The second equalities within Eqs.~(\ref{eq:es}) and (\ref{eq:eo})
above are only correct up to lattice artefacts
which for the lattice actions used here will be proportional to
$a^2\Lambda_{QCD}^2$ and $a^2/r^2$. The $a^2/r^2$ correction means that these
relations do not hold in the limit $r\rightarrow 0$.
In particular, unlike the continuum $E_s$,
$E_{s}^L(0)=0$ does not diverge and
$E_H^L(0)=\Lambda_H^L$. Note that in lattice schemes
both $m_b$ and $V_s$ contain power divergences $\propto a^{-1}$
which cancel each other. As $r<a$ the above continuum limit
interpretation is not valid anymore and $E_{s}(0)=V_{s,L}(0)=0$.
The lattice ``$1/r$'' term within $V_{s,L}$
assumes a finite
value, $\propto a^{-1}$, at $r= 0$, that exactly cancels the
power term of the static energy.

If we subtract the above equations from each other
we obtain a consistent picture,
\begin{equation}
\label{eq:expect}
E_H(r)-E_s(r)=V_s(r)-V_o(r)+\Lambda_H+O(r^2)=E_H^L(r)-E_{s}^L(r).
\end{equation}
The second equality only applies modulo lattice artefacts, i.e.\ for
$r\gg a$. This combination does not contain the leading
renormalon and is also power term free.

As indicated in Fig.~\ref{fig:objects}, in the limit $r\rightarrow 0$
the non-perturbative energy levels calculated on the lattice will
either correspond to gluelumps or to glueballs, at least if we neglect
sea quarks
for the moment. In fact, as gluelumps
are accompanied by power divergencies, at sufficiently small lattice spacing,
gluelumps will always form the ground state within each $J^{PC}$ sector.
Starting from these glueballs 
and pulling the static sources apart
(${\mathbf 1}\oplus {\mathbf 8}={\mathbf 3}\otimes{\mathbf 3}^*$)
these states can be identified as the
ground state in the fundamental open string sector plus glueball scattering
states. If we start from a gluelump, increasing the distance will lead
us to hybrid energies (or hybrid energies plus glueballs).
At $r>0$ the symmetry is reduced from
$O(3)\otimes {\mathcal C}$ to
$D_{\infty h}$ and each gluelump state will in fact be approached
by more than one hybrid energy level.
Note that gluelump masses are scale and scheme dependent while
gluelump mass splittings are universal.

We have restricted the discussion of gluelumps and hybrid potentials
to the case without sea quarks. Note however that the presence of
(massive) sea quarks will not change the situation conceptionally.
We have already discussed the breaking of the ground state
string above and the same will happen in the hybrid sector.
Sea quarks will result in $B\overline{B}$ potentials,
in addition to the conventional hybrid potentials. Their presence
will provide us with an alternative possibility of screening of a
static octet charge by quark and antiquark and this will
give rise to additional scattering states.
The level orderings will somewhat change and the excitation
spectrum will become more dense. However, the pNRQCD multipole expansion also
provides the framework for classifying this situation and moreover
the renormalon and power term structure will remain unaffected.

Finally, the gluelump energy $\Lambda_H$ is an object very similar to
the binding energy $\overline{\Lambda}$ of a $B$ meson.
Imagine a heavy gluino with mass $m_{\tilde{g}}$. This will be screened
by the gluons within the QCD vacuum and only visible as a bound state
glueballino with mass,
\begin{equation}
M_{\tilde{G}}=m_{\tilde{g}}+\Lambda_B+O(1/m_{\tilde{g}}).
\end{equation}
The lightest glueballino will be the magnetic one, $H=1^{+-}=B$.
The scheme and scale dependence of $\Lambda_B$ is required to
cancel that of $m_{\tilde{g}}$.
In a similar way in which heavy-light mesons are related to
quarkonia at $r\rightarrow\infty$, glueballinos are related
to gluinonia, bound states of two heavy gluinos:
\begin{equation}
E_A(r)=2m_{\tilde{g}}+V_{A,s}(r)+O(r^2)\stackrel{r\rightarrow\infty}{\longrightarrow}
2M_{\tilde{G}},
\end{equation}
where $V_{A,s}(r)$ is the singlet potential between two adjoint sources.
As ${\mathbf 8}\otimes{\mathbf 8}={\mathbf 1}\oplus{\mathbf 8}\oplus{\mathbf 8}
\oplus{\mathbf 1}{\mathbf 0}\oplus{\mathbf 1}{\mathbf 0}^*\oplus{\mathbf 2}{\mathbf 7}$, at $r\rightarrow 0$
contact can be made between static hybrid
gluinonium energy levels and gluelumps, this time not only in representation
${\mathbf 8}$ but also in ${\mathbf 10}$ and ${\mathbf 27}$: a whole tower
of states exists, and each new representation introduces a new
renormalon/power term.

Unfortunately, no numerical results on hybrid excitations of non-fundamental
QCD strings or of higher representation gluelumps exist thus far. However,
singlet open string excitations in representations even larger than the
adjoint one have for instance been calculated in ref.~\cite{Bali:2000un}
and the closed string spectrum in
ref.~\cite{Lucini:2004my}.
\section{String Breaking and $m_b$}
\label{sec:large}
Attempts to resolve string breaking in
lattice studies of QCD with sea quarks have a long history.
While everyone knows that this effect
exists, it is still a crucial benchmark for the capability
of lattice calculations. Moreover, the dynamics of the
mixing between open string and broken string states,
which are the starting points for a description of quarkonia
and heavy-light mesons, respectively,
is non-trivial. This will elucidate
strong quarkonium decay rates and other properties near
threshold.

\begin{figure}[hbt]
\epsfig{file=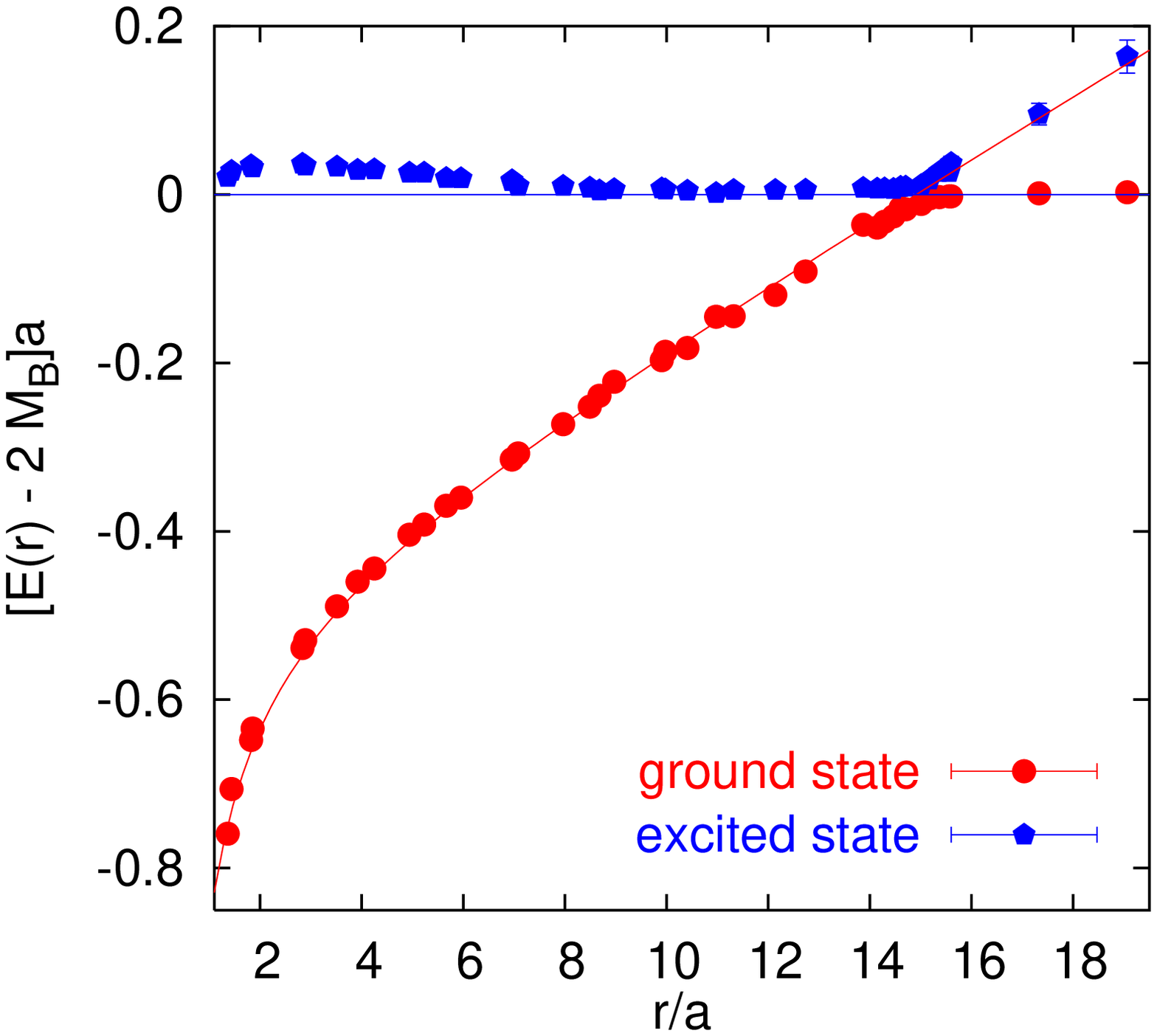,width=.47\columnwidth}~~~
\epsfig{file=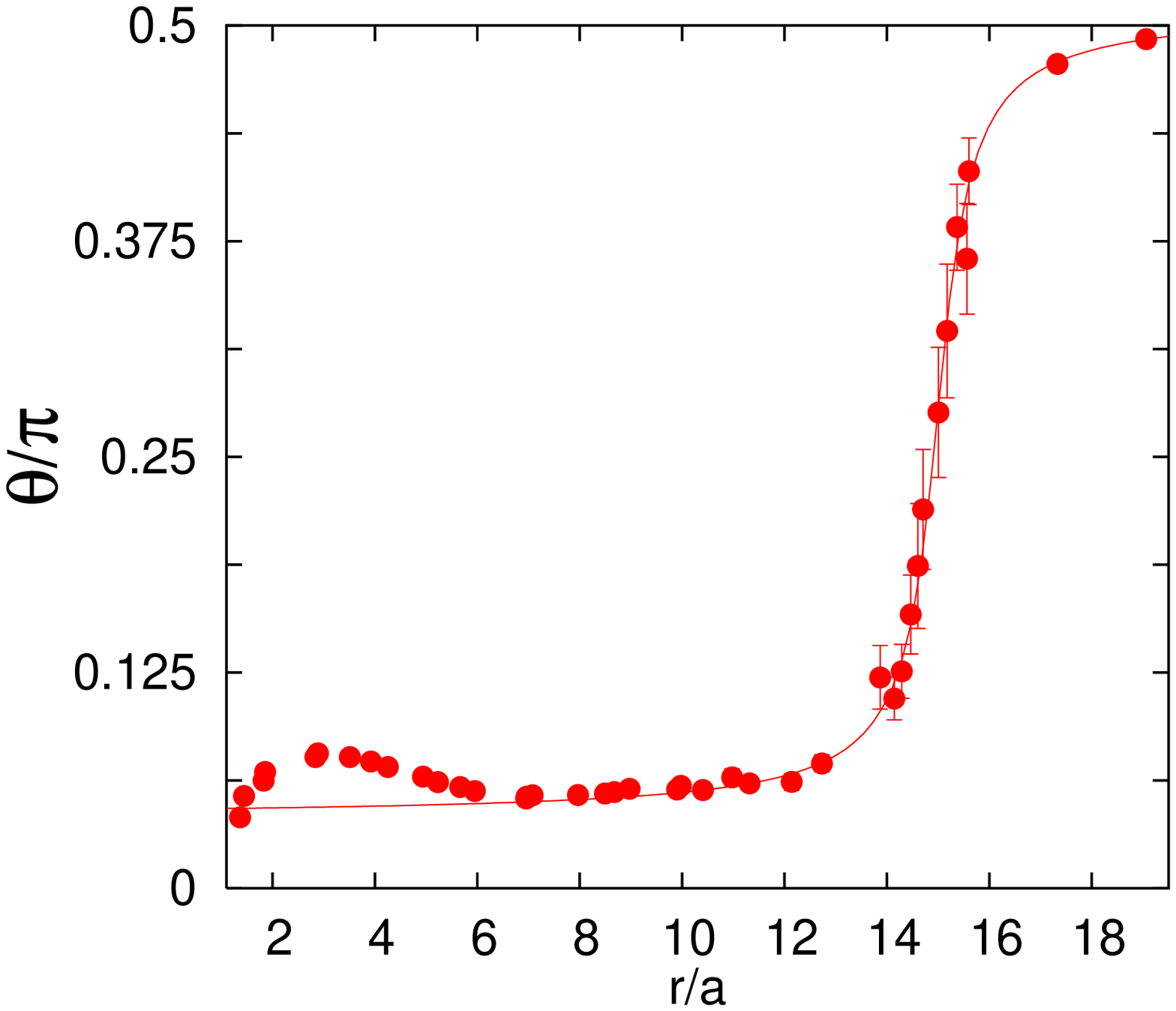,width=.47\columnwidth}
\caption[]{The lowest two energy levels in the $Q{\overline Q}$ system
with sea quarks. On the right hand side the $\overline{B}B$ content
of the ground state is shown, in terms of a mixing angle.
The results apply to $n_f=2$, $m_q\approx
m_s$ and $a\approx 0.085$~fm.}
\label{fig:break}
\end{figure}

First studies in toy models were performed as soon as in
1988~\cite{Jersak:1988bf}. Only some five years ago
quantitatively satisfying results were obtained,
first in $SU(2)$-Higgs models~\cite{Philipsen:1998de,Knechtli:1998gf}
and then for the
adjoint representation string in
$SU(2)$ gauge theory~\cite{Stephenson:1999kh,Philipsen:1999wf}. The latter case
corresponds to the decay of gluinonium into two gluinoballs, discussed above.
Several attempts on the string breaking problem in QCD with sea
quarks were made~\cite{Pennanen:2000yk,bolder,Duncan:2000kr,Bernard:2001tz} but only
very recently reliable results were obtained~\cite{Bali:2004pb}. These
are displayed in Fig.~\ref{fig:break}. String breaking
takes place at around $r_c\approx 1.27$ fm for light quark masses
similar to that of the strange quark.
An extrapolation to physical quark masses yields $r_c\approx 1.16$~fm.
The gap between
the two states in the string breaking region is about $\Delta E\approx 50$~MeV and we are
able to resolve this with a resolution of 10 standard deviations!
One might expect
lighter sea quarks to extend the
pion exchange related ``bump'' in the excited
energy level and mixing angle towards larger distances and also to
further broaden the gap $\Delta E$.

\begin{figure}[hbt]
\centerline{\epsfig{file=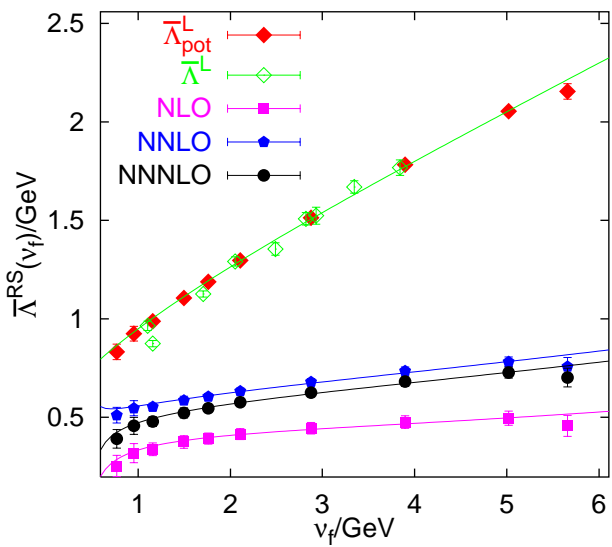,width=.6\columnwidth}}
\caption[]{The binding energy $\overline{\Lambda}$
from the $Q\overline{Q}$ potential
(solid diamonds) and from static-light mesons (open diamonds) as well as
the perturbative conversion into the RS scheme at different orders.}
\label{fig:sl}
\end{figure}
We have discussed that $E_s(r)\rightarrow 2M_B$ for $r\rightarrow \infty$.
This is not the case in the quenched approximation, however,
still at some distance $r_c=1.13(7)$~fm~\cite{Bali:2003jq,bolder}
we will find
$\frac12 E_s(r_c)=M_B$. On the lattice we have,
\begin{equation}
\frac12 E_{s}^L(r_c)=\frac12 E_s(r_c)-m_{b,L}=\overline{\Lambda}^L_{\mbox{\scriptsize pot}}.
\end{equation}
One can determine both, the binding energy from the static potential,
$\overline{\Lambda}^L_{\mbox{\scriptsize pot}}(a)=
\frac{1}{2}E_s^L(0.5\,\mbox{fm};a)+\Delta$,
up to a constant $\Delta$  and
$\overline{\Lambda}^L(a)$ directly from the static-light system~\cite{Duncan:1994uq,Allton:1993ix,Ewing:1995ih,Green:2003zz}.
The latter values are less accurate since
a chiral extrapolation
in the light quark mass is required.
There are also inconsistencies between different
$\overline{\Lambda}^L$ data sets obtained by different groups at the coarser
lattice spacings.

The leading overall renormalon ambiguity
cancels from
the running of $\Lambda^L$ from one scale $\nu_f=a^{-1}$ to another.
In Fig.~\ref{fig:sl} we plot the results (diamonds), together with 
the NNNLO perturbative expectation, expanded in terms of
$\alpha_s(3.9\,\mbox{GeV})$. We find excellent agreement between perturbation
theory and the lattice data, down to energies as low
as $\nu_f<1$~GeV. We also translate the results into the
RS scheme~\cite{staticpot}. At large scales (small lattice spacings)
the power term becomes large and
hence a conversion between schemes requires very accurate perturbative
coefficients. At small scales this requirement is less demanding but
perturbation theory obviously becomes less convergent.
AT NNNLO the optimal accuracy can be obtained around $\nu_f\approx 4$~GeV.

By subtracting the power term from $\overline{\Lambda}^L$
one obtains the binding energy in the OS scheme. This is scale independent
but contains a renormalon ambiguity.
If we are interested in extracting the 
$b$ quark mass in the $\overline{MS}$ scheme
the renormalon above will
cancel against the one that arises
from converting the OS $b$ quark mass
$m_{b,OS}(a)=M_B-\overline{\Lambda}^{OS}\!(a)$ into the $\overline{MS}$
mass~\cite{Gimenez:1996nw}.
We use the RS scheme as an intermediate scheme in this conversion and
obtain~\cite{Bali:2003jq},
\begin{equation}
\label{mMSPV}
m_{b,\overline{MS}}(m_{b,\overline{MS}})
=[4191\pm 29(\mbox{stat.})\pm 47
(\mbox{th.})\pm 1 (\Lambda_{\overline{MS}})]\,\mbox{MeV}.
\end{equation}
The theoretical error includes $1/m_b$ corrections. In addition to
the errors displayed we expect an $O(100\,\mbox{MeV})$ quenching uncertainty.
\begin{figure}[hbt]
\centerline{\epsfig{file=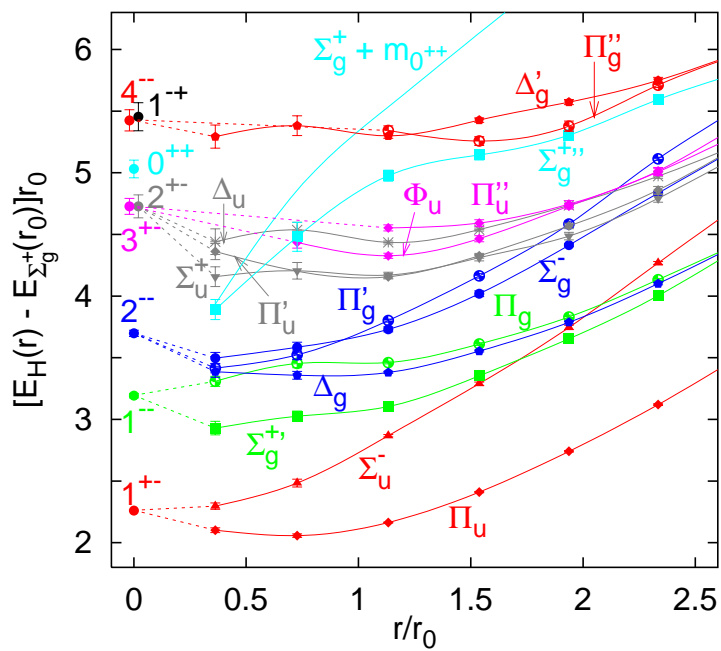,width=.7\columnwidth}}
\caption[]{Hybrid energies~\cite{latticeshort2}, in
comparison with the gluelump spectrum, extrapolated to the continuum
limit~\cite{FM} (circles, left-most data points). The gluelump spectrum
has been shifted by an arbitrary constant to adjust the $1^{+-}$ state
with the $\Pi_u$ and $\Sigma_u^-$ potentials at short distance.
The lines are
drawn to guide the eye.}
\label{fig:hybs}
\end{figure}
\section{Gluelumps and Hybrids}
\label{sec:short}
At short distances one would expect more than one
$\Lambda_{\eta}^{\sigma_v}$ hybrid to approach
the same $J^{PC}$ gluelump energy level where $J\geq\Lambda$
and $\eta=PC$ since $D_{\infty h}\subset O(3)\otimes {\mathcal C}$.
Qualitatively this can be verified in Fig.~\ref{fig:hybs} where we compare
hybrid energies obtained in ref.~\cite{latticeshort2} with the gluelump
spectrum of ref.~\cite{FM}. The dashed lines correspond to our expectations.
Only the $\Sigma_g^{+\prime\prime}$ level cannot be
disentangled from a glueball scattering state. Everything is displayed in
units of $r_0\approx 0.5$~fm.

\begin{figure}[hbt]
\centerline{\epsfig{file=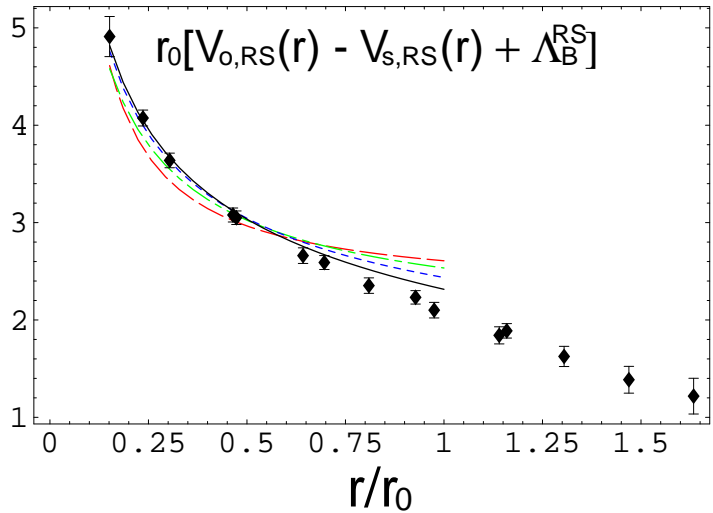,width=.7\columnwidth}}
\caption[]{
Splitting between the $\Pi_u$ and the $\Sigma_g^+$ energies.
$(V_{o,\mbox{\tiny RS}}-V_{s,\mbox{\tiny RS}})(r)+\Lambda_B^{\mbox{\tiny RS}}$
is plotted  at tree level (dashed line), one-loop
(dashed-dotted line), two loops (dotted line) and three loops (estimate)
plus the leading single ultrasoft log (solid line).}
\label{fig:hyb2}
\end{figure}
We calculate the lowest two hybrid potentials $E^L_H(r)$ with
$H=\Pi_u$ and $H=\Sigma_u^-$
which both will approach the lightest (magnetic) $1^{-+}$ gluelump.
According to Eq.~(\ref{eq:eo}), the hybrid energies are the non-perturbative
generalisation of the octet potential of the perturbative multipole expansion.
pNRQCD predicts the leading order difference between these two
levels to be $\propto r^2$, which we are able to verify.

There are now two strategies of determining
a gluelump mass $\Lambda_B$. Either one can directly calculate it at
a given lattice spacing $a$ on the lattice and subsequently convert
it into another scheme or one
can compute the
power term free combination
$E_{\Pi_u}^L(r)-E^L_{s}(r)$ at $r\gg a$ and extrapolate the
result to the continuum limit. Subsequently, one can then follow
Eq.~(\ref{eq:expect}), subtract
$V_s(r)-V_o(r)$ and obtain $\Lambda_H$.
The result of a continuum limit extrapolation of the difference
is displayed in Fig.~\ref{fig:hyb2}, together with the
expectation Eq.~(\ref{eq:expect}) to different orders in perturbation theory,
using the RS scheme~\cite{staticpot,Bali:2003jq}.
The $n_f=0$ QCD coupling has been taken from ref.~\cite{Lambda} and hence the
only free parameter in the fit is the gluelump energy,
\begin{equation}
\Lambda_B^{\mbox{\scriptsize RS}}(1\,\mbox{GeV})=[887\pm 39(\mbox{latt.})\pm 83(\mbox{th.})
\pm 32(
\Lambda_{\overline{MS}})]\,\mbox{MeV}.
\end{equation}

We obtain a compatible result, following the first strategy outlined above,
perturbatively converting the lattice
gluelump data of ref.~\cite{FM} at finite lattice spacings,
$\Lambda_B^L(a)$,
into the RS scheme,
in analogy to our determination
of the static-light binding energy.
The second lightest gluelump is the electric one ($1^{--}$) which is
about 350--400~MeV heavier than the magnetic gluelump.
\section{Conclusions}
Relations between static energy levels at short and large distances
(string breaking) have been reviewed.
For the static energies NNLO/NNNLO $\overline{MS}$ perturbation
theory works very well down to energies of less than 1 GeV, once the
leading renormalon has been accounted for.
Static
Wilson-Schwinger lines in the fundamental
and adjoint representations 
give rise to masses (the binding energies
$\overline{\Lambda}$ and gluelumps, respectively)
that are scale and scheme dependent. This has implications with respect to
QCD vacuum models and condensates.
In particular, when combining perturbative Wilson
coefficients with non-perturbative matrix elements these have to
be defined in the same scheme and at the same scale to enable
renormalon cancellation.

\begin{acknowledge}
I thank my collaborators Thomas D\"ussel, Thomas Lippert,
Hartmut Neff, Antonio
Pineda and Klaus Schilling.
This work is supported by the
EC Hadron Physics I3 Contract No.\ RII3-CT-2004-506078,
by a PPARC Advanced
Fellowship (grant PPA/A/S/2000/00271) as well as by PPARC grant
PPA/G/0/2002/0463. 
\end{acknowledge}

\end{document}